\documentclass[12pt]{article}

\usepackage{graphicx}
\usepackage{amsmath}
\usepackage{amssymb}
\numberwithin{equation}{section}

\newcommand{\diag}{{\rm diag}}
\newcommand{\R}{{\mathcal R}}
\newcommand{\fL}{{\mathfrak L}}
\newcommand{\fR}{{\mathfrak R}}
\newcommand{\fQ}{{\mathfrak Q}}
\newcommand{\fS}{{\mathfrak S}}
\newcommand{\fC}{{\mathfrak C}}
\newcommand{\fI}{{\mathfrak I}}

\newcommand{\alg}[1]{\mathfrak{#1}}

\begin{document}

\baselineskip=16pt plus 0.2pt minus 0.1pt

\begin{titlepage}
\title{
\hfill\parbox{4cm}
{\normalsize MIT-CTP-3843}\\
\vspace{1cm}
{\Large\bf A Yangian Double for the AdS/CFT Classical r-matrix}
}
\author{
{\sc Sanefumi Moriyama}${}^{1,2}$\,{}\thanks
{{\tt moriyama@math.nagoya-u.ac.jp}}\;
\quad and \quad
{\sc Alessandro Torrielli}${}^{1}$\,{}\thanks
{{\tt torriell@mit.edu}}\\[12pt]
${}^{1}${\it Center for Theoretical Physics, 
Massachusetts Institute of Technology,}\\
{\it Cambridge, MA02139, USA}\\[6pt]
${}^{2}${\it Graduate School of Mathematics, Nagoya University,}\\
{\it Nagoya 464-8602, Japan}\\
}
\date{\normalsize June, 2007}
\maketitle
\thispagestyle{empty}

\begin{abstract}
\normalsize
We express the classical r-matrix of AdS/CFT in terms of tensor
products involving an infinite family of generators, which takes a
form suggestive of the universal expression obtained from a Yangian
double. This should provide an insight into the structure of the
infinite dimensional symmetry algebra underlying the integrability of
the model, and give a clue to the construction of its universal
R-matrix. We derive the commutation relations under which the algebra
of these new generators close.
\end{abstract}

\end{titlepage}

\section{Introduction}
Motivated by the AdS/CFT correspondence, integrable structures were
discovered on both the gauge theory side \cite{gauge} and the string
theory side \cite{string} (see \cite{rev} for reviews).
Among the main results has come the derivation of a scattering matrix
\cite{Beisert,nlin} whose tensorial structure turns out to be completely
fixed by the centrally extended $\alg{su(2|2)}\oplus\alg{su(2|2)}$
symmetry of the problem in its fundamental $(2|2)$ matrix
representation. 
The additional scalar (dressing) factor \cite{dress} is constrained by
the Hopf-algebraic analog of crossing symmetry \cite{Janik}.
Remarkable advances in the exact determination of this phase factor
have been recently made \cite{AF,transcend}, as well as progress in
understanding the nature of the infinite dimensional symmetry algebra
underlying the planar integrability of the model
\cite{Yangian,FZ,qpoincare}.
In \cite{BeisYang}, in particular, a Yangian symmetry has been
advocated, and the hope for the existence of a universal R-matrix which
could reproduce the AdS/CFT scattering matrix upon choosing an
appropriate representation, has been put on somewhat firmer grounds.
The importance of having a universal form for the R-matrix, namely, an
expression purely in terms of generators of the corresponding symmetry
algebra independent of the specific representation chosen, comes from
two basic facts.
The first one, is a better insight into the underlying symmetry
responsible for the integrability of the model.
In particular, factorizing the R-matrix explicitly into an infinite sum
of tensor products of abstract generators allows to directly read off
the symmetry algebra.
The operators appearing respectively on the left and on the right hand
side of the tensor product are dual to each other, and form a double
in Drinfeld's sense. 
The second useful fact, is the accessibility to all possible scattering
matrices for any representations with a single expression.
In other words, when taken in the appropriate representation, the
universal R-matrix can reduce to the one for the scattering of a bound
state from an elementary excitation, or between two bound states.
The representations for bound states have been studied in \cite{Dorey}. 

The presence of the Hopf algebra structure found in \cite{hopf}
precisely leads to the expectation of the existence of an infinite
dimensional bialgebra equipped with a triangular structure by the above
R-matrix.
In order to gain a better insight into this construction, it is useful
to study the classical limit of the R-matrix, namely a deformation
around the identity: the so-called classical r-matrix, which is a
solution of the classical Yang-Baxter equation.
The relevance of the classical r-matrix relies on the fact that
traditionally one can reconstruct the full quantum R-matrix from the
information encoded in its classical limit.
This is done by means of powerful theorems \cite{Drin}, which also allow
a classification of the possible symmetry structures arising.
Roughly speaking, this procedure should correspond to some analog of the
exponential map between Lie algebras and Lie groups. 
For a detailed explanation and motivations of the importance of the
classical Yang-Baxter equation, the reader is referred to
\cite{Etingof}.

The program of analyzing the algebraic structure of the classical
r-matrix was initiated in \cite{rclass}, where also the residue at its
simple pole at the origin was computed.
This revealed the appearance of the Casimir of the $\alg{gl(2|2)}$
superalgebra. 
Even though a rigorous application of the above mentioned theorems is
elusive, the properties of the residue suggest that some version of the
standard reconstruction theorems might still work. 
The plan of this paper is precisely to extract from the classical
r-matrix as much information as possible about its universal form.
The idea is to obtain a rewriting, which allows us to read the form of
the symmetry generators directly from the r-matrix.

In order to do that, we will look for what is called Drinfeld's second
realization of Yangians \cite{DrinSec}. 
That is, instead of a realization of the type discussed in
\cite{BeisYang}, given in terms of Lie algebra generators and additional
generators constructed recursively (Drinfeld's first realization), we
will look for a set of generators parametrized by an integer label,
which simultaneously realize the whole Yangian algebra.
These generators are traditionally the ones which are employed for
constructing the universal R-matrix.
For further details about such construction we refer to
\cite{KT}, and to \cite{HSTY} for recent progress along these lines.
The goal of the present paper is to read these generators from a
rewriting of the classical r-matrix in a form suggestive of a Yangian
double, and compute the commutation relations they satisfy, which should
become the defining relations of the desired Yangian algebra.
We remind that only on the double of the Yangian one can have a
well-defined quasi-triangular structure \cite{KT}.

Let us explain the method we will use in a simple example, namely Yang's
classical r-matrix. 
This is a solution of the classical Yang-Baxter equation of the form 
\begin{equation}
\label{Yang}
r=\frac{C}{x_2-x_1}~,
\end{equation}
where $C$ is the Casimir of $\alg{g}$, with $\alg{g}$ being
a Lie algebra\footnote{It is very simple to realize that (\ref{Yang})
is a solution of $[r_{12},r_{13}]+[r_{12},r_{23}]+[r_{13},r_{23}]=0$,
when one remembers that the Casimir operator commutes with the trivial
coproduct of the Lie algebra generators, and $r$ lives in 
$\alg{g}\otimes\alg{g}$.}.  
One can extract information about the generators of the infinite
dimensional symmetry algebra by factorizing it in a geometric sum,
$1/(x_2-x_1)=\sum_{n=0}^\infty x_1^nx_2^{-n-1}$.
If we now express the Casimir in terms of an orthonormal basis 
$C=T^a\otimes T^a$, we can see that $r$ takes the form
\begin{equation}
\label{Yang2}
r = \sum_{n=0}^{\infty} T^a_{n} \otimes T^a_{-n-1}~,
\end{equation}
where the generators $T^a_n=T^ax^n$ are taken in the evaluation 
representation on both factors of the tensor product. 
We refer to \cite{Etingof} for a description of the mathematical
consequences of such construction in the theory of Lie bialgebras.

We would like to follow the same strategy for the present case.
The main difference is that we will have sometimes to expand two
denominators, according to the two poles discussed in \cite{rclass}, one
at $1/(x_2-x_1)$, one at $1/(x_1x_2-1)$.
Nevertheless, we will find that the complicated-looking expressions can
be put into a rather simple and suggestive form:
\begin{align}
&r=\sum_{n=0}^\infty\Bigl(
\fQ^\alpha{}_{a,n}\otimes\widehat\fS^a{}_{\alpha,-n-1}
-\fS^a{}_{\alpha,n}\otimes\widehat\fQ^\alpha{}_{a,-n-1}
+\fC_n\otimes\widehat\fI_{-n-1}+\fI_n\otimes\widehat\fC_{-n-1}
\nonumber\\
&+\bigl(\fR^a{}_{b,n}\otimes\widehat\fR^b{}_{a,-n-1}
-\fR^a{}_{b,-n-1}\otimes\widehat\fR^b{}_{a,n}\bigr)
-\bigl(\fL^\alpha{}_{\beta,n}\otimes\widehat\fL^\beta{}_{\alpha,-n-1}
-\fL^\alpha{}_{\beta,-n-1}\otimes\widehat\fL^\beta{}_{\alpha,n}\bigr)
\Bigr)~,
\label{double}
\end{align}
where the generators will be defined in the main text.
This form is reminiscent of what one expects from the structure of a
Yangian double. 
We will derive these infinite families of generators in a particular
representation, which directly emerges from the classical r-matrix.
We call the subscript $n$ the level of the Yangian.
At level-zero we recover the original Lie superalgebra generators.

The generator $\alg{I}$ appearing in (\ref{double}) is proportional to
$\diag(1,1,-1,-1)$, and extends the Cartan subalgebra of $\alg{sl(2|2)}$
to $\alg{gl(2|2)}$.
Such an extension needs to be introduced on general grounds. 
Whenever, in fact, the Cartan matrix of a Lie superalgebra is degenerate
(as in the present case for $\alg{su(2|2)}$), one needs to introduce an
additional Cartan generator in order to make such matrix non-degenerate,
and be able to invert it.
The inverse of the Cartan matrix extended in this way will appear in
the universal R-matrix, together with the additional Cartan generator
\cite{KT}\footnote{An alternative extension by $\alg{sl(2)}$
automorphisms has been considered in \cite{nlin,Fabthes,BeisYang}.
In particular in \cite{Fabthes} the Casimir operator for such an
extension was considered.}.
In the present case, we follow the discussion in \cite{Gade}:
in that paper, the original $\alg{sl(2|2)}$ Cartan subalgebra consists
of the generators $H_1=\diag(-1,0,-1,0)$, $H_2=\diag(0,1,1,0)$ and 
$H_3=\diag(0,-1,0,-1)$.
One has to introduce an additional $H_4=\diag(-1,0,0,1)$ which completes
the algebra to $\alg{gl(2|2)}$.
Then the extended Cartan matrix reads:
\begin{align}
a=\begin{pmatrix}0&1&0&1\\1&0&-1&0\\0&-1&0&1\\1&0&1&0\end{pmatrix}~.
\end{align}
If we re-express these generators in more familiar notation in terms
of ${\alg{R}}=\diag(1,-1,0,0)$, ${\alg{L}}=\diag(0,0,1,-1)$, 
${\alg{C}}=1/2\,\diag(1,1,1,1)$ and ${\alg{I}}=1/2\,\diag(1,1,-1,-1)$, 
the quadratic form reduces to
\begin{align}
(a^{-1})^{ij}H_iH_j=\frac{1}{2}({\alg{R}}^2-{\alg{L}}^2)
+2{\alg{C}}{\alg{I}}~,
\end{align}
which is reminiscent of the form \eqref{double}.
We will find that the coefficient of $\fI_n$ vanishes for $n=0$.
This is consistent with the result of \cite{rclass}.
There, an analysis of the poles of the classical r-matrix was performed,
and the appearance of the Casimir of the superalgebra $\alg{gl(2|2)}$ in
the residue at $x_1=x_2$ was shown.
One thing to notice is that, nevertheless, the terms of the R-matrix
responsible for the exchange between two bosons and two fermions, 
namely $C_{12}$ and $F_{12}$ in \cite{Beisert}, do not contribute to the
residue at this pole.
Therefore, the residue has an additional symmetry, corresponding to the
{\it trivial} coproduct of the generator $\diag(1,1,-1-1)$, which
enhances the algebra to $\alg{gl(2|2)}$.
However, this symmetry is neither of the full R-matrix, nor of the
classical r-matrix, precisely due to these terms.

The plan of the paper is as follows: 
In section 2 we review the properties of the classical r-matrix,
introducing the conventions needed.
In section 3 we perform our rewriting of the entries of the classical
r-matrix in terms of generators of a Yangian double, starting from the
easier non-diagonal part, and ending with the diagonal one.
The new generators are also introduced, whose commutation relations are
presented in section 4. 
We conclude with comments on the main directions of future development.

\section{Review of the classical r-matrix}
Our starting point for the rewriting \eqref{double} is the classical
r-matrix given in \cite{rclass}.
Though the $\alg{su(1|2)}$ basis was adopted there, it turned out to be
the so-called string basis \cite{FZ} that makes direct contact with the
string theory computation.
Here we would like to briefly review the classical r-matrix in the
string basis.
The R-matrix $\R=\Pi\circ{\cal S}$ is constructed from the graded
permutation $\Pi$ and the S-matrix ${\cal S}$ given in
\cite{Beisert,nlin} (see also the comments in \cite{BeisYang}):
\begin{align}
\R_{12}|\phi^a_1\phi^b_2\rangle
&=\frac{1}{2}(A_{12}-B_{12})\frac{U_1}{U_2}|\phi^a_1\phi^b_2\rangle
+\frac{1}{2}(A_{12}+B_{12})\frac{U_1}{U_2}|\phi^b_1\phi^a_2\rangle
+\frac{1}{2}C_{12}U_1\epsilon^{ab}\epsilon_{\alpha\beta}
|\psi^\alpha_1\psi^\beta_2\rangle~,\nonumber\\
\R_{12}|\psi^\alpha_1\psi^\beta_2\rangle
&=-\frac{1}{2}(D_{12}-E_{12})|\psi^\alpha_1\psi^\beta_2\rangle
-\frac{1}{2}(D_{12}+E_{12})|\psi^\beta_1\psi^\alpha_2\rangle
-\frac{1}{2}F_{12}\frac{1}{U_2}\epsilon^{\alpha\beta}\epsilon_{ab}
|\phi^a_1\phi^b_2\rangle~,\nonumber\\
\R_{12}|\phi^a_1\psi^\beta_2\rangle
&=G_{12}\frac{1}{U_2}|\phi^a_1\psi^\beta_2\rangle
+H_{12}|\psi^\beta_1\phi^a_2\rangle~,\nonumber\\
\R_{12}|\psi^\alpha_1\phi^b_2\rangle
&=L_{12}U_1|\psi^\alpha_1\phi^b_2\rangle
+K_{12}\frac{U_1}{U_2}|\phi^b_1\psi^\alpha_2\rangle~,
\end{align}
with $U=\sqrt{x^+/x^-}$ and $A_{12},B_{12},\ldots$ given by
\begin{align}
&A_{12}=\frac{x_2^+-x_1^-}{x_2^--x_1^+}~,\quad
B_{12}=\frac{x_2^+-x_1^-}{x_2^--x_1^+}
\biggl(1-2\frac{1-g^2/2x_1^+x_2^-}{1-g^2/2x_1^-x_2^-}
\frac{x_2^+-x_1^+}{x_2^+-x_1^-}\biggr)~,\nonumber\\
&\qquad C_{12}=\frac{g^2\gamma_1\gamma_2}{\alpha x_1^+x_2^+}
\frac{1}{1-g^2/2x_1^-x_2^-}
\frac{x_2^+-x_1^+}{x_2^--x_1^+}~,\nonumber\\
&D_{12}=-1~,\quad
E_{12}=-\biggl(1-2\frac{1-g^2/2x_1^-x_2^+}{1-g^2/2x_1^+x_2^+}
\frac{x_2^--x_1^-}{x_2^--x_1^+}\biggr)~,\nonumber\\
&\qquad F_{12}=-\frac{2\alpha(x_1^+-x_1^-)(x_2^+-x_2^-)}
{\gamma_1\gamma_2 x_1^-x_2^-}
\frac{1}{1-g^2/2x_1^+x_2^+}
\frac{x_2^--x_1^-}{x_2^--x_1^+}~,\nonumber\\
&G_{12}=\frac{x_2^+-x_1^+}{x_2^--x_1^+}~,\quad
H_{12}=\frac{\gamma_1}{\gamma_2}
\frac{x_2^+-x_2^-}{x_2^--x_1^+}~,\nonumber\\
&L_{12}=\frac{x_2^--x_1^-}{x_2^--x_1^+}~,\quad
K_{12}=\frac{\gamma_2}{\gamma_1}
\frac{x_1^+-x_1^-}{x_2^--x_1^+}~.
\end{align}
We adopt the parametrization of \cite{AF} for the variables $x^\pm$:
\begin{equation}
\label{AF}
x^\pm(x)=\frac{x}{2\zeta}
\bigg(\sqrt{1-\frac{\zeta^2}{(x-x^{-1})^2}} 
\pm i\frac{\zeta}{x-x^{-1}}\bigg)~, 
\end{equation}
and we take $\zeta=1/(\sqrt{2}g)$ ($=2\pi/\sqrt{\lambda}$ in terms of
the {}'t Hooft coupling constant $\lambda=g_{\rm YM}^2N$) as a
deformation parameter, namely expand all formulas around $\zeta=0$
keeping $x$ fixed.
This corresponds to the near BMN limit \cite{BMN}.
The classical r-matrix is defined by the infinitesimal deviation from
unity of the R-matrix:
\begin{align}
{\cal{R}}_{12}=1+i\zeta r_{12}~.
\end{align}
After some computations we find that it is given by
\begin{align}
r_{12}|\phi^a_1\phi^b_2\rangle
&=\frac{(x_1^2+x_2^2)(x_1^2x_2^2+1)-4x_1^2x_2^2}
{(x_2-x_1)(x_1x_2-1)(x_1^2-1)(x_2^2-1)}
|\phi^a_1\phi^b_2\rangle
+\frac{2x_1x_2}{(x_2-x_1)(x_1x_2-1)}
|\phi^b_1\phi^a_2\rangle\nonumber\\
&\qquad+\frac{\gamma_1\gamma_2}{i\zeta\alpha}\frac{1}{x_1x_2-1}
\epsilon^{ab}\epsilon_{\alpha\beta}|\psi^\alpha_1\psi^\beta_2\rangle~,
\label{rbb}\\
r_{12}|\psi^\alpha_1\psi^\beta_2\rangle
&=\frac{2x_1x_2}{(x_2-x_1)(x_1x_2-1)}
|\psi^\alpha_1\psi^\beta_2\rangle
-\frac{2x_1x_2}{(x_2-x_1)(x_1x_2-1)}
|\psi^\beta_1\psi^\alpha_2\rangle\nonumber\\
&\qquad-\frac{4\zeta\alpha}{i\gamma_1\gamma_2}
\frac{x_1^2x_2^2}{(x_1^2-1)(x_2^2-1)(x_1x_2-1)}
\epsilon^{\alpha\beta}\epsilon_{ab}|\phi^a_1\phi^b_2\rangle~,
\label{rff}\\
r_{12}|\phi^a_1\psi^\beta_2\rangle
&=\frac{x_2(x_2+x_1)}{(x_2-x_1)(x_2^2-1)}
|\phi^a_1\psi^\beta_2\rangle
+\frac{\gamma_1}{\gamma_2}\frac{2x_2^2}{(x_2-x_1)(x_2^2-1)}
|\psi^\beta_1\phi^a_2\rangle~,
\label{rbf}\\
r_{12}|\psi^\alpha_1\phi^b_2\rangle
&=\frac{x_1(x_2+x_1)}{(x_2-x_1)(x_1^2-1)}
|\psi^\alpha_1\phi^b_2\rangle
+\frac{\gamma_2}{\gamma_1}\frac{2x_1^2}{(x_2-x_1)(x_1^2-1)}
|\phi^b_1\psi^\alpha_2\rangle~.
\label{rfb}
\end{align}

We would like to rewrite this classical r-matrix in terms of the 
$\alg{su(2|2)}$ generators $\fR^a{}_b$, $\fL^\alpha{}_\beta$,
$\fQ^\alpha{}_a$, $\fS^a{}_\alpha$ with the central element $\fC$, and
their infinite Yangian partners labeled by an integer $n$, whose
fundamental representation for $n=0$ is given by
\begin{align}
&\fR^a{}_b|\phi^c\rangle
=\delta^c_b|\phi^a\rangle-\frac{1}{2}\delta^a_b|\phi^c\rangle~,
\label{Rfund}\\
&\fL^\alpha{}_\beta|\psi^\gamma\rangle
=\delta^\gamma_\beta|\psi^\alpha\rangle
-\frac{1}{2}\delta^\alpha_\beta|\psi^\gamma\rangle~,
\label{Lfund}
\end{align}
and
\begin{align}
&\fQ^\alpha{}_a|\phi^b\rangle=a\delta^b_a|\psi^\alpha\rangle~,\quad
\fQ^\alpha{}_a|\psi^\beta\rangle
=b\epsilon^{\alpha\beta}\epsilon_{ab}|\phi^b\rangle~,
\label{Qfund}\\
&\fS^a{}_\alpha|\phi^b\rangle
=c\epsilon^{ab}\epsilon_{\alpha\beta}|\psi^\beta\rangle~,\quad
\fS^a{}_\alpha|\psi^\beta\rangle=d\delta^\beta_\alpha|\phi^a\rangle~,
\label{Sfund}
\end{align}
as well as
\begin{align}
\fC|\phi^a\rangle=C|\phi^a\rangle~,\quad
\fC|\psi^\alpha\rangle=C|\psi^\alpha\rangle~,
\end{align}
with $a,b,c,d$ defined by the limit $\zeta\to 0$ of the corresponding
variables introduces in \cite{Beisert} (see also \cite{rclass})
\begin{align}
a=\gamma~,\quad
b=\frac{2\zeta\alpha}{i\gamma}\frac{x}{x^2-1}~,\quad
c=\frac{i\gamma}{2\zeta\alpha}\frac{1}{x}~,\quad
d=\frac{1}{\gamma}\frac{x^2}{x^2-1}~.
\label{abcd}
\end{align}
It is understood that $\alpha$ scales as $\zeta^{-1}$ and $\gamma$ 
is of order $1$ in the classical limit.
An additional operator $\fI$ whose eigenvalue vanishes for $n=0$ will
be introduced later. 

\section{Classical r-matrix as a Yangian double}
After reviewing the expression for the classical r-matrix in the
previous section, here we would like to embark on our project of
rewriting the classical r-matrix \eqref{rbb}--\eqref{rfb} in terms of
generators as in \eqref{double}.
Since each sector is independent, we shall start with the easier
off-diagonal sector, and then turn to the diagonal sector.

\subsection{Fermionic sector}
First, let us concentrate on the combination of two fermionic
generators, which only affect the last terms in
\eqref{rbb}--\eqref{rfb}.
For this purpose we note that half of the coefficients from each term
can be expressed as
\begin{align}
\frac{\gamma_1\gamma_2}{i2\zeta\alpha}\frac{1}{x_1x_2-1}
=\sum_{n=0}^\infty a_1x_1^n\cdot c_2x_2^{n+1}
=-\sum_{n=0}^\infty c_1x_1^{-n}\cdot a_2x_2^{-n-1}~,\nonumber\\
-\frac{2\zeta\alpha}{i\gamma_1\gamma_2}
\frac{x_1^2x_2^2}{(x_1^2-1)(x_2^2-1)(x_1x_2-1)}
=-\sum_{n=0}^\infty b_1x_1^{-n}\cdot d_2x_2^{-n-1}
=\sum_{n=0}^\infty d_1x_1^n\cdot b_2x_2^{n+1}~,\nonumber\\
\frac{\gamma_1}{\gamma_2}\frac{x_2^2}{(x_2-x_1)(x_2^2-1)}
=\sum_{n=0}^\infty a_1x_1^n\cdot d_2x_2^{-n-1}
=-\sum_{n=0}^\infty c_1x_1^{-n}\cdot b_2x_2^{n+1}~,\nonumber\\
\frac{\gamma_2}{\gamma_1}\frac{x_1^2}{(x_2-x_1)(x_1^2-1)}
=-\sum_{n=0}^\infty b_1x_1^{-n}\cdot c_2x_2^{n+1}
=\sum_{n=0}^\infty d_1x_1^n\cdot a_2x_2^{-n-1}~,
\end{align}
in an appropriate domain of convergence.
Since the action of two fermionic generators on two excitations goes as
\begin{align}
\fQ^\alpha{}_a\otimes\fS^a{}_\alpha|\phi^b_1\phi^c_2\rangle
=a_1c_2\epsilon^{bc}\epsilon_{\beta\gamma}
|\psi^\beta_1\psi^\gamma_2\rangle~,\quad
&\fS^a{}_\alpha\otimes\fQ^\alpha{}_a|\phi^b_1\phi^c_2\rangle
=c_1a_2\epsilon^{bc}\epsilon_{\beta\gamma}
|\psi^\beta_1\psi^\gamma_2\rangle~,\nonumber\\
\fQ^\alpha{}_a\otimes\fS^a{}_\alpha|\psi^\beta_1\psi^\gamma_2\rangle
=-b_1d_2\epsilon^{\beta\gamma}\epsilon_{bc}
|\phi^b_1\phi^c_2\rangle~,\quad
&\fS^a{}_\alpha\otimes\fQ^\alpha{}_a|\psi^\beta_1\psi^\gamma_2\rangle
=-d_1b_2\epsilon^{\beta\gamma}\epsilon_{bc}
|\phi^b_1\phi^c_2\rangle~,\nonumber\\
\fQ^\alpha{}_a\otimes\fS^a{}_\alpha|\phi^b_1\psi^\gamma_2\rangle
=a_1d_2|\psi^\gamma_1\phi^b_2\rangle~,\quad
&\fS^a{}_\alpha\otimes\fQ^\alpha{}_a|\phi^b_1\psi^\gamma_2\rangle
=c_1b_2|\psi^\gamma_1\phi^b_2\rangle~,\nonumber\\
\fQ^\alpha{}_a\otimes\fS^a{}_\alpha|\psi^\beta_1\phi^c_2\rangle
=-b_1c_2|\phi^c_1\psi^\beta_2\rangle~,\quad
&\fS^a{}_\alpha\otimes\fQ^\alpha{}_a|\psi^\beta_1\phi^c_2\rangle
=-d_1a_2|\phi^c_1\psi^\beta_2\rangle~,
\end{align}
we find that in the fermionic sector the classical r-matrix can be
expressed as
\begin{align}
r\Big|_{\fQ\fS}=\sum_{n=0}^\infty
\Bigl(\fQ^\alpha{}_{a,n}\otimes\widehat\fS^a{}_{\alpha,-n-1}
-\fS^a{}_{\alpha,n}\otimes\widehat\fQ^\alpha{}_{a,-n-1}\Bigr)~,
\label{rf}
\end{align}
with $\fQ^\alpha{}_{a,n}$, $\widehat\fQ^\alpha{}_{a,n}$,
$\fS^a{}_{\alpha,n}$ and $\widehat\fS^a{}_{\alpha,n}$ defined by
\begin{align}
&\fQ^\alpha{}_{a,n}=\widehat\fQ^\alpha{}_{a,n}
=\fQ^\alpha{}_a\bigl(x^n\Pi_{\rm b}+x^{-n}\Pi_{\rm f}\bigr)~,\label{Q}\\
&\fS^a{}_{\alpha,n}=\widehat\fS^a{}_{\alpha,n}
=\fS^a{}_\alpha\bigl(x^{-n}\Pi_{\rm b}+x^n\Pi_{\rm f}\bigr)~.\label{S}
\end{align}
The operators $\Pi_{\rm b}$ and $\Pi_{\rm f}$ are projectors in the
bosonic and fermionic subspaces respectively, namely in matrix notation
$\Pi_{\rm b}={\rm diag}(1,1,0,0)$ and 
$\Pi_{\rm f}={\rm diag}(0,0,1,1)$.
Note that both operators without hats $\fQ^\alpha{}_{a,n}$,
$\fS^a{}_{\alpha,n}$ and operators with hats
$\widehat\fQ^\alpha{}_{a,n}$, $\widehat\fS^a{}_{\alpha,n}$ have the same
expressions, though the operators without hats are only defined for
$n\ge 0$ while the operators with hats are only defined for $n<0$.
Note also that the expression in \eqref{rf} should be regarded as a
formal series.
After acting on states, we interpret the summation as an analytical
continuation from the result obtained in an appropriate domain of
convergence. 
Equivalently, one could act on the r-matrix with the operator
$(D_\rho\otimes 1)$ \cite{Gade}, where the operator $D_\rho$ multiplies
any generator at level $n$ by the representation-independent parameter
$\rho^{|n|}$, perform the series in a domain of $\rho$ where one has
convergence, and analytically continue to $\rho=1$ at the end.

\subsection{Bosonic off-diagonal sector}
Now let us turn to the bosonic off-diagonal sector.
We would like to rewrite the second terms in \eqref{rbb} and \eqref{rff}
into the operator doubles.
Here the coefficient can be rewritten as
\begin{align}
\frac{2x_1x_2}{(x_2-x_1)(x_1x_2-1)}=\sum_{n=0}^\infty
\Bigl([n+1]_{x_1}[n+2]_{x_2}-[n+2]_{x_1}[n+1]_{x_2}\Bigr)~,
\end{align}
because of
\begin{align}
&\frac{x_1x_2}{(x_2-x_1)(x_1x_2-1)}
=\sum_{n=0}^\infty[n+1]_{x_1}x_2^{n+1}
=\sum_{n=0}^\infty[n+1]_{x_1}x_2^{-n-1}\nonumber\\
&\qquad=-\sum_{n=0}^\infty x_1^{n+1}[n+1]_{x_2}
=-\sum_{n=0}^\infty x_1^{-n-1}[n+1]_{x_2}~.\label{sum}
\end{align}
Here we have introduced a $q$-number $[n]_q$ by
\begin{align}
[n]_q=\frac{q^n-q^{-n}}{q-q^{-1}}~.
\end{align}
Since the action of the operators $\fR^a{}_b\otimes\fR^b{}_a$ and
$\fL^\alpha{}_\beta\otimes\fL^\beta{}_\alpha$ takes the form
\begin{align}
\fR^a{}_b\otimes\fR^b{}_a|\phi^c_1\phi^d_2\rangle
=|\phi^d_1\phi^c_2\rangle-\frac{1}{2}|\phi^c_1\phi^d_2\rangle~,\quad
\fL^\alpha{}_\beta\otimes\fL^\beta{}_\alpha
|\psi^\gamma_1\psi^\delta_2\rangle
=|\psi^\delta_1\psi^\gamma_2\rangle
-\frac{1}{2}|\psi^\gamma_1\psi^\delta_2\rangle~,
\label{RRLL}
\end{align}
we can rewrite the classical r-matrix in this sector as
\begin{align}
r\Big|_{\fR\fL}
&=\sum_{n=0}^\infty\Bigl(
\bigl(\fR^a{}_{b,n}\otimes\widehat\fR^b{}_{a,-n-1}
-\fR^a{}_{b,-n-1}\otimes\widehat\fR^b{}_{a,n}\bigr)\nonumber\\
&\qquad-\bigl(\fL^\alpha{}_{\beta,n}
\otimes\widehat\fL^\beta{}_{\alpha,-n-1}
-\fL^\alpha{}_{\beta,-n-1}
\otimes\widehat\fL^\beta{}_{\alpha,n}\bigr)
\Bigr)~,\label{rl}
\end{align}
with $\fR^a{}_{b,n}$, $\widehat\fR^a{}_{b,n}$, $\fL^\alpha{}_{\beta,n}$
and $\widehat\fL^\alpha{}_{\beta,n}$ defined by
\begin{align}
&\fR^a{}_{b,n}=[n+1]_x\fR^a{}_b~,\quad
\widehat\fR^a{}_{b,n}=-[n-1]_x\fR^a{}_b~,\label{R}\\
&\fL^\alpha{}_{\beta,n}=[n+1]_x\fL^\alpha{}_\beta~,\quad
\widehat\fL^\alpha{}_{\beta,n}=-[n-1]_x\fL^\alpha{}_\beta~.\label{L}
\end{align}
In this case, all of $\fR^a{}_{b,n}$, $\widehat\fR^a{}_{b,n}$,
$\fL^\alpha{}_{\beta,n}$ and $\widehat\fL^\alpha{}_{\beta,n}$ are
defined for both $n\ge 0$ and $n<0$.
Note that there are ambiguities in this rewriting because of various
expressions in \eqref{sum}.
Our current choice is partially motivated by its rather symmetric
form, and partially by the closure of the commutation relations, which
will be the subject of our next section.

\subsection{Bosonic diagonal sector}
Finally let us consider the diagonal sector.
We assume that the Cartan subalgebra of $\alg{su(2)}\times\alg{su(2)}$
(generators $\alg{R}^a{}_b$ and $\alg{L}^\alpha{}_\beta$) is already
taken care of in \eqref{rl} by suitably completing the set of indices
contracted.
This is simply due to $\alg{su(2)}\times\alg{su(2)}$ covariance of the
string basis.
The remaining diagonal part we would like to rewrite into the form of a
Yangian double is therefore
\begin{align}
r_{12}|\phi^a_1\phi^b_2\rangle
&=\biggl[
\frac{(x_1^2+x_2^2)(x_1^2x_2^2+1)-4x_1^2x_2^2}
{(x_2-x_1)(x_1x_2-1)(x_1^2-1)(x_2^2-1)}
+\frac{x_1x_2}{(x_2-x_1)(x_1x_2-1)}
\biggr]|\phi^a_1\phi^b_2\rangle
+\cdots~,\nonumber\\
r_{12}|\psi^\alpha_1\psi^\beta_2\rangle
&=\biggl[
\frac{2x_1x_2}{(x_2-x_1)(x_1x_2-1)}
-\frac{x_1x_2}{(x_2-x_1)(x_1x_2-1)}
\biggr]|\psi^\alpha_1\psi^\beta_2\rangle
+\cdots~,\nonumber\\
r_{12}|\phi^a_1\psi^\beta_2\rangle
&=\frac{x_2(x_2+x_1)}{(x_2-x_1)(x_2^2-1)}
|\phi^a_1\psi^\beta_2\rangle
+\cdots~,\nonumber\\
r_{12}|\psi^\alpha_1\phi^b_2\rangle
&=\frac{x_1(x_2+x_1)}{(x_2-x_1)(x_1^2-1)}
|\psi^\alpha_1\phi^b_2\rangle
+\cdots~.\label{rd}
\end{align}
The extra term in the squared parentheses comes from rearranging the
operator action as \eqref{RRLL}. 
Note that the phase of the S-matrix was undetermined in \cite{Beisert}.
Hence the diagonal sector has the ambiguity of an overall shift,
corresponding to an overall scalar factor at the level of the full
quantum R-matrix.
In this paper, we will still freely add and subtract such terms when
needed, but they should later be determined by some generalized crossing
symmetry emerging from the construction.

Inspired by the argument in the introduction, we would like to make use
of the following generators
\begin{align}
&\fC_n|\phi^a\rangle=C_n|\phi^a\rangle~,\quad
\fC_n|\psi^\alpha\rangle=C_n|\psi^\alpha\rangle~,\\
&\fI_n|\phi^a\rangle=I_n|\phi^a\rangle~,\quad
\fI_n|\psi^\alpha\rangle=-I_n|\psi^\alpha\rangle~.
\end{align}
If we assume the classical r-matrix can be expressed as
\begin{align}
r\Big|_{\fC\fI}
=\sum_{n=0}^\infty\Bigl(\fC_n\otimes\widehat\fI_{-n-1}
+\fI_n\otimes\widehat\fC_{-n-1}\Bigr)~,
\end{align}
this means we have to match \eqref{rd} to
\begin{align}
r_{12}|\phi^a_1\phi^b_2\rangle&=\sum
(C_n\widehat I_{-n-1}+I_n\widehat C_{-n-1})
|\phi^a_1\phi^b_2\rangle+\cdots~,\nonumber\\
r_{12}|\psi^\alpha_1\psi^\beta_2\rangle&=\sum
(-C_n\widehat I_{-n-1}-I_n\widehat C_{-n-1})
|\psi^\alpha_1\psi^\beta_2\rangle+\cdots~,\nonumber\\
r_{12}|\phi^a_1\psi^\beta_2\rangle&=\sum
(-C_n\widehat I_{-n-1}+I_n\widehat C_{-n-1})
|\phi^a_1\psi^\beta_2\rangle+\cdots~,\nonumber\\
r_{12}|\psi^\alpha_1\phi^b_2\rangle&=\sum
(C_n\widehat I_{-n-1}-I_n\widehat C_{-n-1})
|\psi^\alpha_1\phi^b_2\rangle+\cdots~,
\label{candi}
\end{align}
up to an overall shift.
Here the first factors $C_n$ or $I_n$ are understood in the
representation labeled by $x_1$ while the second factors
$\widehat I_{-n-1}$ or $\widehat C_{-n-1}$ are in the $x_2$ one.
For this to be possible, we need a rather non-trivial identity:
\begin{align}
&\frac{(x_1^2+x_2^2)(x_1^2x_2^2+1)-4x_1^2x_2^2}
{(x_2-x_1)(x_1x_2-1)(x_1^2-1)(x_2^2-1)}
+\frac{2x_1x_2}{(x_2-x_1)(x_1x_2-1)}\nonumber\\
&\qquad=\frac{x_2(x_2+x_1)}{(x_2-x_1)(x_2^2-1)}
+\frac{x_1(x_2+x_1)}{(x_2-x_1)(x_1^2-1)}~.
\end{align}
In fact, this identity holds!
Subtracting half of the above quantity to normalize the classical
r-matrix properly, we find\footnote{It would be interesting to
understand the relation between this subtraction and the dressing
factor.}
\begin{align}
r_{12}|\phi^a_1\phi^b_2\rangle
&=\frac{(x_1^2+x_2^2)(x_1^2x_2^2+1)-4x_1^2x_2^2}
{2(x_2-x_1)(x_1x_2-1)(x_1^2-1)(x_2^2-1)}
|\phi^a_1\phi^b_2\rangle
+\cdots~,
\nonumber\\
r_{12}|\psi^\alpha_1\psi^\beta_2\rangle
&=-\frac{(x_1^2+x_2^2)(x_1^2x_2^2+1)-4x_1^2x_2^2}
{2(x_2-x_1)(x_1x_2-1)(x_1^2-1)(x_2^2-1)}
|\psi^\alpha_1\psi^\beta_2\rangle
+\cdots~,
\nonumber\\
r_{12}|\phi^a_1\psi^\beta_2\rangle
&=-\frac{(x_1x_2+1)(x_2+x_1)}{2(x_1^2-1)(x_2^2-1)}
|\phi^a_1\psi^\beta_2\rangle
+\cdots~,
\nonumber\\
r_{12}|\psi^\alpha_1\phi^b_2\rangle
&=\frac{(x_1x_2+1)(x_2+x_1)}{2(x_1^2-1)(x_2^2-1)}
|\psi^\alpha_1\phi^b_2\rangle
+\cdots~,
\end{align}
with
\begin{align}
&\sum_{n=0}^\infty C_n\widehat I_{-n-1}
=\frac{x_1^2(x_2^2-1)}{2(x_2-x_1)(x_1x_2-1)(x_1^2-1)}~,\\
&\sum_{n=0}^\infty I_n\widehat C_{-n-1}
=\frac{x_2^2(x_1^2-1)}{2(x_2-x_1)(x_1x_2-1)(x_2^2-1)}~.
\end{align}
Choosing
\begin{align}
&C_n=\widehat C_n=\frac{x^{n+1}+x^{-n-1}}{2(x-x^{-1})}~,\label{C}\\
&I_n=\widehat I_n=\frac{1}{2}(x^n-x^{-n})~,\label{I}
\end{align}
we find that formula \eqref{candi} holds.
Again, there are ambiguities in rescaling $C_n$ and $I_n$.
Our definition is motivated by the commutation relations in the next
section.
Note that $I_0$ vanishes identically, which is expected from the
argument in the introduction.

\section{Commutation relations}
In the previous section we have rewritten the classical r-matrix in
terms of generators of a tentative Yangian double.
In the process, we have defined level-$n$ operators
$\fQ^\alpha{}_{a,n}$, $\fS^a{}_{\alpha,n}$, $\fR^a{}_{b,n}$,
$\fL^\alpha{}_{\beta,n}$, $\fC_n$ (and $\fI_n$) and their duals, which
reduce to the original $\alg{su(2|2)}$ generators $\fQ^\alpha{}_a$,
$\fS^a{}_\alpha$, $\fR^a{}_b$, $\fL^\alpha{}_\beta$, $\fC$
at level-zero.
Here we would like to investigate their commutation relations.

Originally the operators $\fQ^\alpha{}_{a,n}$, $\fS^a{}_{\alpha,n}$,
$\fC_n$ and $\fI_n$ are defined only for $n\ge 0$ while their duals
are defined only for $n<0$.
Since both these operators and their duals share the same expressions as
can be seen in \eqref{Q}, \eqref{S}, \eqref{C} and \eqref{I}, let us
combine the formula by extending their definition for $n<0$.
On the other hand, the duals of the operators $\fR^a{}_{b,n}$ \eqref{R}
and $\fL^\alpha{}_{\beta,n}$ \eqref{L} can be obtained from the original
operators by substituting $n$ with $-n$.
We will not consider them in the commutation relations.
To summarize, we would like to study the  commutation relations between
the operators $\fQ^\alpha{}_{a,n}$, $\fS^a{}_{\alpha,n}$,
$\fR^a{}_{b,n}$, $\fL^\alpha{}_{\beta,n}$, $\fC_n$ and $\fI_n$, where
the indices run over positive and negative integers.

We would like to remark that at the present stage it is difficult to
exclude that the following commutation relations could be accidental to
our representation, and need to be modified later.
In particular, it is impossible from \eqref{R}, \eqref{L}, \eqref{C}
and \eqref{I} to distinguish between $\fR^a{}_{b,n}$ and
$-\fR^a{}_{b,-n-2}$, between $\fL^\alpha{}_{\beta,n}$ and
$-\fL^\alpha{}_{\beta,-n-2}$, between $\fC_n$ and $\fC_{-n-2}$ and
between $\fI_n$ and $-\fI_{-n}$.
Here we have chosen to present them in the most compact form as we
could find.

Acting the operators $\{\fQ^\alpha{}_{a,m},\fS^b{}_{\beta,n}\}$ on the
bosonic state $|\phi^c\rangle$ and the fermionic state
$|\psi^\gamma\rangle$ respectively using \eqref{Qfund} and \eqref{Sfund}
and reinterpreting the result as the action of a single bosonic operator
with the help of \eqref{Rfund} and \eqref{Lfund}, we find
\begin{align}
\{\fQ^\alpha{}_{a,m},\fS^b{}_{\beta,n}\}
=\delta^b_a\fL^\alpha{}_{\beta,m+n}
+\delta^\alpha_\beta\fR^b{}_{a,m+n}
+\delta^b_a\delta^\alpha_\beta\fC_{m+n}~.
\end{align}
This is the higher level analogue of the commutation relation:
\begin{align}
\{\fQ^\alpha{}_a,\fS^b{}_\beta\}
=\delta^b_a\fL^\alpha{}_\beta+\delta^\alpha_\beta\fR^b{}_a
+\delta^b_a\delta^\alpha_\beta\fC~.
\end{align}
This result justifies our definition of $\fR^a{}_{b,n}$,
$\fL^\alpha{}_{\beta,n}$ and $\fC_n$ in \eqref{R}, \eqref{L} and
\eqref{C}.
Similarly, we find
\begin{align}
\{\fQ^\alpha{}_{a,m},\fQ^\beta{}_{b,n}\}
&=\frac{2\zeta\alpha}{i}\Bigl[
\epsilon^{\alpha\beta}\epsilon_{ab}\fC_{m-n-1}
+\epsilon^{\alpha\beta}\epsilon_{c\{a}\fR^c{}_{b\},m-n-1}
+\epsilon_{ab}\epsilon^{\gamma\{\alpha}\fL^{\beta\}}{}_{\gamma,m-n-1}
\Bigr]~,
\label{centrext1}
\end{align}
\begin{align}
\{\fS^a{}_{\alpha,m},\fS^b{}_{\beta,n}\}
&=\frac{i}{2\zeta\alpha}\Bigl[
\epsilon^{ab}\epsilon_{\alpha\beta}\fC_{m-n-1}
+\epsilon_{\alpha\beta}\epsilon^{c\{a}\fR^{b\}}{}_{c,m-n-1}
+\epsilon^{ab}\epsilon_{\gamma\{\alpha}\fL^{\gamma}{}_{\beta\},m-n-1}
\Bigr]~,
\label{centrext2}
\end{align}
where parentheses enclosing indices denote symmetrization (dividing by
two).
In the computation the following formula is useful.
\begin{align}
\epsilon^{ab}\epsilon_{cd}
=\delta^a_c\delta^b_d-\delta^a_d\delta^b_c~,\quad
\delta^d_a\epsilon_{bc}+\delta^d_b\epsilon_{ca}
+\delta^d_c\epsilon_{ab}=0~.
\label{centrext}
\end{align}
At level zero, one recovers from \eqref{centrext1}, \eqref{centrext2},
\eqref{C} and \eqref{abcd} the two central extensions 
${\mathfrak P}=ab$ and ${\mathfrak K}=cd$ of the superalgebra
$\alg{su(2|2)}$.
The reader might find unpleasant the appearance of the factor
$2\zeta\alpha$ in the commutation relations.
We can always get rid of it by rescaling the generators by
$\fQ^\alpha{}_{a,n}\to\fQ^\alpha{}_{a,n}/\sqrt{2\zeta\alpha}$ and
$\fS^a{}_{\alpha,n}\to\fS^a{}_{\alpha,n}\sqrt{2\zeta\alpha}$.

The commutation relations between one bosonic and one fermionic operator
read
\begin{align}
&[\fR^a{}_{b,m},\fQ^\gamma{}_{c,n}]
={\rm sign}(m+1)
\sum_{l=-|m+1|+1}^{|m+1|-1}\hspace{-5mm}{}'\hspace{5mm}
\Bigl(-\delta^a_c\fQ^\gamma{}_{b,l+n}
+\frac{1}{2}\delta^a_b\fQ^\gamma{}_{c,l+n}\Bigr)~,\\
&[\fR^a{}_{b,m},\fS^c{}_{\gamma,n}]
={\rm sign}(m+1)
\sum_{l=-|m+1|+1}^{|m+1|-1}\hspace{-5mm}{}'\hspace{5mm}
\Bigl(\delta^c_b\fS^a{}_{\gamma,l+n}
-\frac{1}{2}\delta^a_b\fS^c{}_{\gamma,l+n}\Bigr)~,\\
&[\fL^\alpha{}_{\beta,m},\fQ^\gamma{}_{c,n}]
={\rm sign}(m+1)
\sum_{l=-|m+1|+1}^{|m+1|-1}\hspace{-5mm}{}'\hspace{5mm}
\Bigl(\delta^\gamma_\beta\fQ^\alpha{}_{c,l+n}
-\frac{1}{2}\delta^\alpha_\beta\fQ^\gamma{}_{c,l+n}\Bigr)~,\\
&[\fL^\alpha{}_{\beta,m},\fS^c{}_{\gamma,n}]
={\rm sign}(m+1)
\sum_{l=-|m+1|+1}^{|m+1|-1}\hspace{-5mm}{}'\hspace{5mm}
\Bigl(-\delta^\alpha_\gamma\fS^c{}_{\beta,l+n}
+\frac{1}{2}\delta^\alpha_\beta\fS^c{}_{\gamma,l+n}\Bigr)~.
\end{align}
Here we have to expand the $q$-number $[m+1]_x$ attached to the
bosonic operators $\fR^a{}_{b,m}$ and $\fL^\alpha{}_{\beta,m}$ by
\begin{align}
[m]_x={\rm sign}(m)\sum_{l=-|m|+1}^{|m|-1}
\hspace{-3mm}{}'\hspace{3mm}x^l~,
\end{align}
because only monomials are attached to the fermionic operators
$\fQ^\gamma{}_{c,n}$ and $\fS^c{}_{\gamma,n}$.
Note that ${\rm sign}(n)$ is defined to be $1,0,-1$ for $n>0,n=0,n<0$
respectively, and the prime $'$ in the summation symbol $\sum'$
indicates that the summation is taken by steps of two.

Now let us turn to the commutation relations between two bosonic
operators.
\begin{align}
&[\fR^a{}_{b,m},\fR^c{}_{d,n}]
={\rm sign}(m+1)(n+1)
\sum_{l=||m+1|-|n+1||}^{|m+1|+|n+1|-2}\hspace{-7mm}{}'\hspace{7mm}
\Bigl(\delta^c_b\fR^a{}_{d,l}-\delta^a_d\fR^c{}_{b,l}\Bigr)~,\\
&[\fL^\alpha{}_{\beta,m},\fL^\gamma{}_{\delta,n}]
={\rm sign}(m+1)(n+1)
\sum_{l=||m+1|-|n+1||}^{|m+1|+|n+1|-2}\hspace{-7mm}{}'\hspace{7mm}
\Bigl(\delta^\gamma_\beta\fL^\alpha{}_{\delta,l}
-\delta^\alpha_\delta\fL^\gamma{}_{\beta,l}\Bigr)~.
\end{align}
Here we have to expand the product of two $q$-numbers in terms of the
following summation of $q$-numbers:
\begin{align}
[m]_x[n]_x={\rm sign}(mn)
\sum_{l=||m|-|n||+1}^{|m|+|n|-1}\hspace{-5mm}{}'\hspace{5mm}[l]_x~.
\end{align}

Finally the commutation relations between the fermionic operators 
$\fQ^\alpha{}_{a,m}$ and $\fS^a{}_{\alpha,m}$ and our parity operator
$\fI_n$ are non-trivial:
\begin{align}
&[\fQ^\alpha{}_{a,m},\fI_n]
=\fQ^\alpha{}_{a,m+n}-\fQ^\alpha{}_{a,m-n}~,\\
&[\fS^a{}_{\alpha,m},\fI_n]
=-\fS^a{}_{\alpha,m+n}+\fS^a{}_{\alpha,m-n}~.
\end{align}

\section{Conclusions}

We have expressed the classical r-matrix of the AdS/CFT correspondence
in terms of a Yangian double, or an infinite series of tensor products
of operators.
We have also studied the commutation relations among these new
generators.
We hope our result will clarify the underlying symmetry, and give a
clue to the construction of the universal R-matrix of the model.

We shall list some of the main future directions to prosecute our
work.
\begin{itemize}
\item 
The most important development will be to obtain along these lines a
universal expression for the full quantum R-matrix.
We believe that the formula in this paper can be rather suggestive of
the appropriate completion, but a full derivation is still to be
worked out.
\item 
The appropriate coproduct and Hopf algebra structure have to be
defined for the generators we constructed, in order to study the
infinite dimensional symmetry of the R-matrix.
This is traditionally presented in the Chevalley basis, rather
than in the Cartan-Weyl one.
\item 
One should make contact with Beisert's formulation of the Yangian 
symmetry given in Drinfeld's first realization in \cite{BeisYang}, and 
show the relationship with the one presented in this paper.
\item 
The question whether our expression \eqref{double} is truly
``universal'' can also be addressed by studying the double structure 
of the classical r-matrix for the bound states \cite{Dorey}.
We would like to see whether the classical r-matrix for the bound
states can also be rewritten as the same Yangian double satisfying the
same algebra.
\item
So far the main results on the integrable structure of the dilatation 
operator in the Super Yang-Mills theory are restricted to the sector
of the single trace operators or the single string states. 
It would be interesting if this integrable structure can be lifted to
multi-trace operators or multi-string states.
The correspondence between the symmetry generators of matrix string
theory (gauge theory) and those of light-cone string field theory on
the flat space (string theory) given recently in \cite{lcsft} may give a
clue to this question.
\end{itemize}

\section*{Acknowledgments}
We thank P.~Etingof for enlightening discussions. 
We would also like to thank T.~Matsumoto and F.~Spill for many 
interesting discussions
and helpful email exchange.
This work is supported in part by funds provided by the U.S. 
Department of Energy (D.O.E.) under cooperative research agreement
DE-FG02-05ER41360.
The work of S.M. is supported partly by Nishina Memorial Foundation,
Inamori Foundation and Grant-in-Aid for Young Scientists (\#18740143)
from the Japan Ministry of Education, Culture, Sports, Science and
Technology.
A.T. thanks Istituto Nazionale di Fisica Nucleare (I.N.F.N.) for
supporting him through a ``Bruno Rossi'' postdoctoral fellowship.

\end{document}